\documentclass[aps,prl,twocolumn,superscriptaddress]{revtex4}

\usepackage{bm}  
\usepackage{graphicx}
\usepackage{amsmath}
\usepackage{amssymb}
\usepackage{color}
\usepackage{amscd}
\usepackage{epstopdf}

\newcommand{\beq}{\begin{equation}}
\newcommand{\eeq}{\end{equation}}
\newcommand{\beqa}{\begin{eqnarray}}
\newcommand{\eeqa}{\end{eqnarray}}
\newcommand{\bem}{\begin{math}}
\newcommand{\eem}{\end{math}}

\begin{document}
\title{Complex Spontaneous Flows and Concentration Banding in Active Polar Films}
\author{Luca Giomi}
\affiliation{Physics Department and Syracuse Biomaterials Institute, Syracuse University, Syracuse, NY
13244, USA}
\author{M. Cristina Marchetti}
\affiliation{Physics Department and Syracuse Biomaterials Institute, Syracuse University, Syracuse, NY
13244, USA}
\author{Tanniemola B. Liverpool} 
\affiliation{Department of Mathematics, University of Bristol, Clifton, Bristol BS8 1TW, U.K.}
\date{\today}

\begin{abstract}
We study the dynamical properties of  active polar liquid crystalline films. Like active nematic films, active polar films undergo a dynamical transitions to spontaneously flowing steady-states. Spontaneous flow in polar fluids is, however, always accompanied by strong concentration inhomogeneities or ``banding'' not seen in nematics. In addition, a spectacular property unique to polar active films is their ability to generate spontaneously oscillating and  banded flows even at low activity. The oscillatory flows  become increasingly complicated for strong polarity.
\end{abstract}

\maketitle

Active materials are a new class of soft materials maintained out of equilibrium by internal energy sources. There are many examples in biological contexts, including bacterial colonies~\cite{Dombrowski04}, purified extracts of cytoskeletal filaments and motor proteins~\cite{Nedelec03}, and the cell cytoskeleton~\cite{HowardBook}. A non-biological example is a  layer of vibrated granular rods \cite{vibrods}. The key property that distinguishes active matter from more familiar non-equilibrium systems, such as a fluid under shear, is that the energy input that maintains the system out of equilibrium comes from each constituent, rather than the boundaries. Each active particle consumes and dissipates energy going through a cycle that fuels internal changes, generally leading to motion. The experimental systems studied to date typically consists of elongated  active particles of two types: polar particles, with a head and a tail, and apolar ones that are head-tail symmetric.
Active suspensions can then exist in various liquid crystalline states, with novel structural and rheological properties~\cite{SimhaRamaswamySP02,Hatwalne04}.  Apolar particles can form phases with nematic order, characterized by a macroscopic axis of mean orientation identify by a unit vector ${\bf n}$ and global symmetry for ${\bf n}\rightarrow -{\bf n}$, as in equilibrium nematic liquid crystals. Polar particles can order in both nematic and polar phases. The polar phase is again characterized by a mean orientation axis ${\bf p}$, but ${\bf p}\not= -{\bf p}$. 
The protein filaments which are the major component  of cell extracts are generally polar and these extracts can therefore have both nematic and polar phases.

Conventional liquid crystals exhibit a rich non-equilibrium behavior when subject to external forcing, such as shear or applied magnetic and electric fields. This includes transitions to stable statically distorted deformations of the director field (Freedericksz transition~\cite{DeGennes}), shear banding~\cite{Olmsted08}, and even the onset of turbulent and chaotic behavior in the presence of shear~\cite{Buddho04}. \emph{Active} liquid crystals exhibit a similar wealth of phenomena  due to \emph{internal} forcing, i.e., spontaneously. A striking property of active \emph{nematic } liquid crystal films is the onset of spontaneous flow above a critical film thickness first identified by Voituriez \emph{et al}~\cite{Voituriez06}. This phenomenon is  analogous to the Freedericksz transition of a passive nematic in an applied magnetic field, but the flowing state is  driven  by the internal activity of the system - hence the name. This prediction was obtained by analytical studies of the phenomenological hydrodynamic equations of an an active nematic film in a one-dimensional geometry. More recently, Marenduzzo \emph{et al}~\cite{Marenduzzo07} have employed hybrid lattice-Boltzmann simulations to study the active nematic  hydrodynamics in both 1D and 2D geometry and have mapped out the phenomenon in parameter space.

In this letter we show that  active \emph{polar} fluids exhibit an even richer behavior. First,  like active nematics,  polarized active liquid crystals exhibit  steady spontaneous flow. Unlike active nematics,  however, where the filament concentration remains practically uniform in the spontaneously flowing state, spontaneous flow in polar fluids is  accompanied by  ``concentration banding'', i.e., a sharp gradient in the concentration of filaments across the film.  The concentration banding is due to active couplings of concentration gradients to the polar director in the hydrodynamic equations that are allowed only in fluids with polar symmetry. Upon increasing the magnitude of these {\em polar} couplings, the steady state becomes unstable and the system undergoes a further transition to a dynamic state with bands of oscillating concentration and orientation. In the oscillatory regime, travelling bands nucleate and oscillate from one end of the film to the other. For even larger couplings the oscillatory behavior becomes increasingly complex with the appearance of multiple frequencies with incommensurate ratios between the  periods of the orientational and concentration oscillations. 

Hydrodynamic equations for a two component active suspension have been written down phenomenologically~\cite{Joanny_twofluids07} and  derived from a microscopic model~\cite{TBLMCMbook}. The relevant hydrodynamic variables are  the concentration $c$ of filaments and the total density $\rho$ and momentum density ${\bf g}=\rho{\bf v}$ of the suspension, with ${\bf v}$ the flow velocity. We consider an incompressible film, with $\rho={\rm constant}$ and  $\bm\nabla\cdot{\bf v}=0$, and  macroscopic dimensionless polarity ${\bf P}=|{\bf P}|{\bf p}$, with direction characterized by a unit vector ${\bf p}=(\cos\theta,\sin\theta)$, the polar director.   

The hydrodynamic equations for two-component {\em polar} suspensions have been derived elsewhere by coarse-graining a microscopic model of the dynamics of interacting motors and filaments~ \cite{TBLMCMbook}. The active suspension is an intrinsically non-equilibrium system and cannot be described by a free energy. However, for  clarity of presentation we introduce here the equations phenomenologically and write all equilibrium-like terms (i.e., those terms that are also present in an equilibrium polar suspension) in terms of derivatives of a non-equilibrium analogue of a  free energy, given by~\cite{WKMCMKS06}
\begin{multline}\label{F}
 F=\int_{\bf r} \left\{
\frac{C}{2}\left(\frac{\delta c}{c_0}\right)^2+\frac{a_2}{2}|{\bf P}|^2+\frac{a_4}{4}|{\bf P}|^4\right.\\
\left.+\frac{K_1}{2}\left(\bm\nabla\cdot{\bf P}\right)^2
+\frac{K_3}{2}\left(\bm\nabla\times{\bf P}\right)^2\right.\\
\left.+B_1\frac{\delta c}{c_0} \bm\nabla\cdot{\bf P}
+\frac{B_2}{2}|{\bf P}|^2 \bm\nabla\cdot{\bf P}+\frac{B_3}{3c_0}|{\bf P}|^2{\bf P}\cdot\bm\nabla c\right\}\,,
 \end{multline}
with $C$ the compressional modulus and  $K_1$ and $K_3$ the splay and bend elastic constants, both taken equal to $K$ below. The last three terms on the right hand side of Eq. \eqref{F} couple concentration  and splay and are present in equilibrium polar suspensions ($B_1=B_2=B_3=B$ below). The dynamics of the concentration and of the polar director is described by 
\begin{subequations}
\beq\label{conc-eq}
\partial_{t}c=-\bm\nabla\cdot\left[c({\bf v}-\beta' c\ell^2{\bf P})+\Gamma'{\bf h}-\Gamma''\bm\nabla\Big(\frac{\delta F}{\delta c}\Big)\,\right] ,
\eeq
\begin{multline}
\label{p-eq}
\Big[\partial_t+({\bf v}+\beta c \ell^2{\bf P})\cdot\bm\nabla\Big]P_i+\omega_{ij}P_j\\
=\lambda u_{ij}P_j+\Gamma h_i-\Gamma'\partial_i\Big(\frac{\delta F}{\delta c}\Big)\,,
\end{multline}
\end{subequations}
where $\ell$ is the length of the filaments, $u_{ij}=\frac12(\partial_iv_j+\partial_jv_i)$ and $\omega_{ij}=\frac12(\partial_iv_j-\partial_jv_i)$ are the rate-of-strain and vorticity tensors, and ${\bf h}=-\delta F/{\delta \bf P}$  the molecular field. 
\begin{figure}[t]
\centering
\includegraphics[width=1\columnwidth]{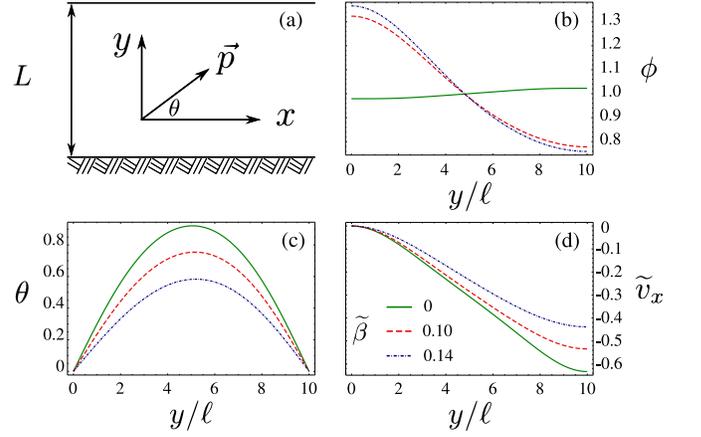}
\caption{\label{fig:polar}(Color online) (a) Sketch of the film geometry. (b-d) Solutions of Eqs. \eqref{eq:phi}, \eqref{eq:theta} for $\lambda=0.1$, $\xi=0.3$, $D=1$, $\widetilde{C}=0.5$, $\eta\Gamma=0.5$, $w=0.13$, $\widetilde{\alpha}=0.08$ and variable $\widetilde{\beta}$.}
\end{figure}
Here $\Gamma$, $\Gamma'$ and $\Gamma''$ are kinetic coefficients and $\beta$ and $\beta'$ are active parameters. The equation describing momentum conservation is written in the Stokes approximation as $\partial_j\sigma_{ij}=0$. The stress tensor $\sigma_{ij}$ is the sum of reversible, dissipative and active contributions, $\sigma_{ij}=\sigma_{ij}^r+\sigma_{ij}^d+\sigma_{ij}^a$. The reversible part is  written in an equilibrium-like form,
\begin{equation}
 \sigma_{ij}^r=-\delta_{ij}\Pi-\frac{\lambda}{2}\big[P_ih_j+P_jh_i\big]
 -\overline{\lambda}\delta_{ij}{\bf P}\cdot{\bf h}+\frac12\big[P_ih_j-P_jh_i\big]\, ,\nonumber \label{sigmar}
\end{equation}
with $\Pi$ the pressure and $\lambda$ and $\overline{\lambda}$ alignment parameters. The dissipative part of the stress tensor is written  as $\sigma_{ij}^d=\eta u_{ij}$ assuming 
a single viscosity $\eta$. Finally, there are additional stresses induced by activity given by
\beqa
\sigma_{ij}^a=\frac{c^2\ell^2}{\Gamma}\Big[-\delta_{ij}\Pi^{a}+\alpha P_iP_j+\ell^2\beta''(\partial_iP_j+\partial_jP_i)\Big]\;.\nonumber 
\eeqa
with $\Pi^a$ the active part of the pressure.
There are  two contributions to the active stress tensor. The first ($\sim\alpha$) describes active stresses  that arise from contractile (if $\alpha>0$) forces induced by  activity. This term is present in both nematic and polar liquid crystals and its effects have been studied before. The second term ($\sim\beta''$) arises from ``self-propulsion'' of the active units and is exclusive to polar systems. The same mechanism is also responsible for the ``convective'' terms  proportional to $\beta$ and $\beta'$ in Eqs.~(\ref{conc-eq}) and (\ref{p-eq}). For a motor/filament mixture, all active contributions are proportional to the mean rate $\Delta\mu$ of ATP consumption, which is the internal driving force for the system.  $\beta$-type terms have dimensions of velocity and have been estimated in the microscopic model as $\beta\sim \widetilde{m}u_0$, with $\widetilde{m}$ a dimensionless concentration of motor clusters and $u_0\sim \Delta\mu$  the mean velocity at which motor clusters step along the filaments~\cite{AATBLMCM06}.  Previous work on confined active films~\cite{Voituriez06,Marenduzzo07} has been limited to active nematics with all the $\beta$ terms equal to zero. Here for the first time we incorporate the polar active terms and analyze their role in controlling non-equilibrium effects in active films. Hereafter we will assume $\beta'=\beta''=\beta$.
\begin{figure}[t]
\centering
\includegraphics[width=1\columnwidth]{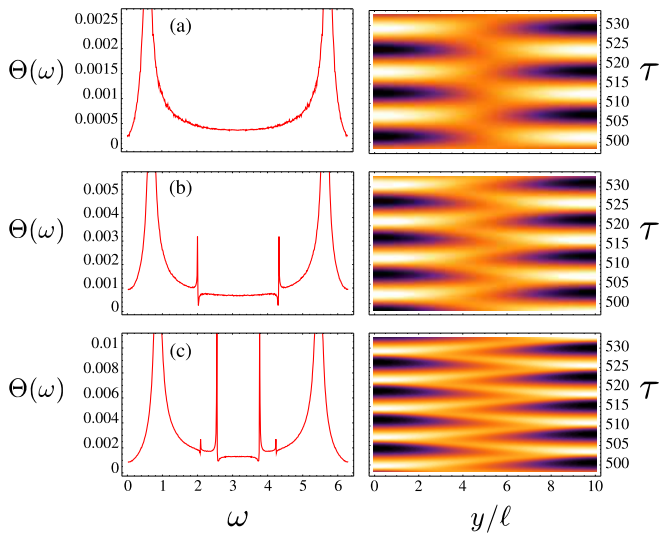}
\caption{\label{fig:space-time}(Color online) On the left discrete Fourier transforms of  $\theta(z\ell=L/2,\tau)$ (center of the film)  for $\lambda=0.1$, $\xi=0.3$, $D=1$, $\widetilde{C}=0.3$, $\eta\Gamma=0.5$, $w=0.13$, $\widetilde{\alpha}=0.1$ and $\widetilde{\beta}=15$ (a) $18$ (b) and $20$ (c). On the right space-time plots of $\phi(z,\tau)$.}
\end{figure}

We consider a two-dimensional active polar suspension with polarization of uniform magnitude and discuss the dynamics of the hydrodynamic fields $c$, ${\bf v}$ and ${\bf p}$. For simplicity we set $|{\bf P}|=1$. The film sits on a solid plane at $y=0$ and is bound by a free surface at $y=L$ (Fig.~\ref{fig:polar}a). The discussion below is easily extended to other boundary conditions.  The film extends to infinity in the $x$ direction and we assume translational invariance along $x$. The Stokes equation requires $\partial_y\sigma_{yy}=0$ and $\partial_y\sigma_{xy}=0$. The first of these two conditions fixes the pressure in the film. The second, together with the boundary condition $\sigma_{xy}(L)=0$, requires $\sigma_{xy}={\rm constant}=0$ throughout the film. We also assume no-slip boundary conditions at the substrate, so that $v_x(0)=0$. 

It is convenient to work with dimensionless quantities by introducing the  time scale $\tau_{0}=\ell^{2}/(\Gamma K)$. Letting $z=y/\ell$, $\tau=t/\tau_{0}$, $\phi=c/c_{0}$ and specializing Eqs. \eqref{conc-eq} and \eqref{p-eq} to our quasi-one-dimensional geometry, we obtain
\begin{subequations}
\begin{multline}\label{eq:phi}
\partial_{\tau}\phi
=\partial_{z}\Bigl\{\widetilde{\beta}\phi^{2}\sin\theta
+\lambda\widetilde{u}\sin\theta\sin2\theta\\
+\left[D(1-\xi\sin^{2}\theta)-w\cos^{2}\theta\right]\partial_{z}\phi\Bigr\}\,,
\end{multline}\\[-26pt]
\begin{multline}\label{eq:theta}
\partial_{\tau}\theta 
=(1-w\cos^{2}\theta)\partial_{z}^{2}\theta
+\tfrac{1}{2}w\sin 2\theta(\partial_{z}\theta)^{2}\\
-\phi \widetilde{\beta}\sin\theta \partial_{z}\theta
+w\cos\theta\partial_{z}\phi-\widetilde{u}(1-\lambda\cos2\theta)\,,
\end{multline}
\end{subequations}
where $\widetilde{\beta}=\beta c_{0}\ell\tau_{0}$, $w=2\ell B/K$, $\xi=\Gamma'^{2}/(\Gamma\Gamma'')$, and $D=\Gamma''C/(c_{0}^{2}\Gamma K)$. The dimensionless rate-of-strain $\widetilde{u}=2 u_{xy}\tau_{0}$
can be obtained from the condition $\sigma_{xy}=0$,
\begin{gather}\label{eq:u}
\widetilde{u} =
- \frac{1}{2\eta\Gamma+\lambda^{2}\sin^{2}2\theta}\Big\{
  \lambda w \sin2\theta\sin^{2}\theta(\partial_{z}\theta)^{2}\notag\\[2pt]
+ \left[w(1-\lambda\cos2\theta)-2\lambda \widetilde{C}\sin^2\theta\right]\cos\theta \partial_{z}\phi \notag\\[4pt]
- 2\widetilde{\beta}c_{0}\ell^{2}\phi^{2}\sin\theta\partial_{z}\theta
+ \widetilde{\alpha}\phi^{2}\sin2\theta\Big\}\,,
\end{gather}
where $\widetilde{\alpha}=\alpha c_{0}^{2}\ell^{4}/(\Gamma K)$ and $\widetilde{C}=\ell\Gamma'C/(\Gamma K)^{2}$. 
The hydrodynamic equations for a two component active nematic suspension are obtained from the above by letting $\tilde{\beta}=0$ and $w=0$. The terms proportional to $w$  are also present in passive polar fluids as they arise from the fact that the polar symmetry allows the coupling proportional to $B$ between splay and density fluctuations. The terms proportional to $\widetilde{\beta}$ are intrinsically non-equilibrium polar terms. Finally, the case of an incompressible  \emph{one-component nematic} fluid, investigated by Voituriez \emph{et al}~\cite{Voituriez06} and by Marenduzzo \emph{et al}~\cite{Marenduzzo07}, can be recovered from our equations by setting $\widetilde{\beta}=w=0$ and assuming a constant concentration $\phi$. 

Eqs. \eqref{eq:theta} and \eqref{eq:phi} are integrated numerically with boundary conditions $\theta(0,\tau)=\theta(L/\ell,\tau)=0$, $\partial_z\phi(0,\tau)=\partial_z\phi(L/\ell,\tau)=0$ (i.e. $j_y(0,t)=j_y(L,t)=0$).  The initial conditions on $\theta$ and $\phi$ are chosen as random, with the constraint $\langle \theta(z,0) \rangle=0$ and $\langle \phi(z,0) \rangle=1$ where $\langle\,\cdot\,\rangle$ stands for a spatial average.

\paragraph{Steady spontaneous flow:}
Both the polar and the apolar systems exhibit a Freedericksz-like transition between a state where the director field is constant and parallel to the walls throughout the channel to a non-uniformly oriented state in which the system spontaneously flows in the $x$-direction. The transition can be tuned by changing either the film thickness or the activity parameter $\widetilde{\alpha}$. Fig~\ref{fig:polar} shows a numerical solution of Eqs. \eqref{eq:phi} and \eqref{eq:theta} for  fixed 
$\widetilde{\alpha}$ and variable $\widetilde{\beta}$, with $L/\ell=10$. As the active velocity $\widetilde{\beta}$ is increased, the maximum tilt $\theta_{m}$ decreases and the alignment is progressively restored. Remarkably the variation in the concentration $\phi$ across the film is significantly stronger than in the apolar case (solid green curve in Fig.~\ref{fig:polar}) with a relative difference between the highest and the lowest values up to 50\%. This ``concentration banding'' is a characteristic of polar active systems. It is a consequence of the active $\beta'$ coupling  in Eq. (\ref{conc-eq}) resulting from self-propelled `convection' of the active elements along the local polarization direction. The varying local polarization angle required for spontaneous flow therefore leads to an even stronger variation in the local concentration. Close to the transition, $\widetilde{\alpha}_c(\widetilde \beta)$, the coupling between the polar director and concentration also leads to an asymmetric director profile across the film. We also point out that in the absence of this polar active term, there are equilibrium-like gradient couplings between local director and density. These are, however, much weaker since they occur at higher order in gradients. In contrast, the active nematic shows a negligible concentration gradient even for anomalously large values of the contractile activity parameter,  $\widetilde{\alpha} \gg \widetilde{\alpha}_c(\widetilde \beta)$.  

\begin{figure}[t]
\centering
\includegraphics[width=0.8\columnwidth]{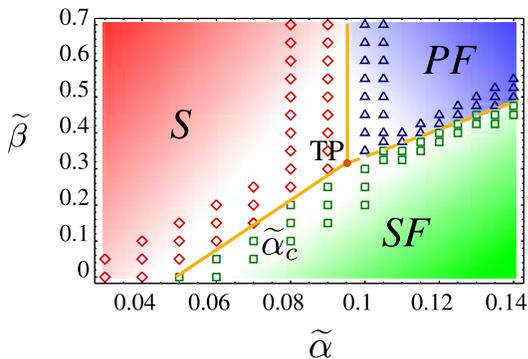}
\caption{\label{fig:alpha-beta}(Color online) Phase-diagram in the $(\widetilde{\alpha},\widetilde{\beta})-$plane. The points  have been obtained numerically using $\lambda=0.1$, $\xi=0.3$, $D=1$, $\widetilde{C}=0.3$, $\eta\Gamma=0.5$, $w=0.13$ and $L/\ell=10$. For $\widetilde{\beta}<\widetilde{\beta}_{\mathrm{TP}}$ upon increasing $\widetilde{\alpha}$ the system undergoes a transition between stationary homogeneous state (S) and inhomogeneous steady flow (SF). Above the ``tricritical point'' (TP) the spontaneous flow becomes oscillatory (PF).}
\end{figure}

\paragraph{Spontaneous oscillations:}
Upon further increasing  $\widetilde{\beta}$,  spontaneous oscillations of $\phi(z,\tau)$ and $\theta(z,\tau)$ are obtained.  The coupled dynamics  of the two fields  gives rise to travelling waves of  concentration and orientation bands. Initially only one frequency  is observed, but the oscillations become more complicated as $\widetilde{\beta}$ increases. Using Fourier decomposition we find that this is due to the appearance of additional  frequencies at different values of $\widetilde{\beta}$ for concentration and orientation bands (see Fig \ref{fig:space-time}). A phase diagram in the $(\widetilde{\alpha},\widetilde{\beta})-$ plane is displayed in Fig. \ref{fig:alpha-beta}. It shows transitions between stationary (S) flow, spontaneous steady flow (SF) and spontaneous periodic (oscillatory) flow (PF). 

We can understand the phase behavior close to a stationary homogeneous state ($\phi=\phi_0$, $\theta=0$, $\tilde{u}=0$) by  expanding $\theta(z) =\delta \theta_+(z) e^{i \omega t}+ \delta \theta_-(z) e^{-i \omega t}$ and  $\phi(z) =\phi_0 + \delta \phi_+(z) e^{i \omega t}+\delta \phi_-(z) e^{-i \omega t}$.
The boundary conditions require $\phi_\pm (z) = \sum_{n=1}^\infty a^\pm_n \cos \left( { n \pi \ell z/L}\right)$ and $\theta_\pm (z) = \sum_{n=1}^\infty b^\pm_n \sin \left( { n \pi \ell z/L}\right)$. The first unstable modes are  $a^\pm_1,b^\pm_1$ and a linear stability analysis shows that there is a steady-state ($\omega=0$) instability  at 
\begin{equation}\label{eq:alpha_c}
\widetilde{\alpha}_{c}(\widetilde \beta) 
=\frac{\eta\Gamma(1-w)}{1-\lambda}\left({\pi\ell \over \phi_{0}L}\right)^{2}
+ \frac{w\widetilde{\beta}\,[\eta\Gamma+\tfrac{1}{2}(1-\lambda)^{2}]}{(1-\lambda)(D-w)}\,,
\end{equation}
to a steady spontaneously flowing state with concentration banding.
Oscillatory modes with frequency $\omega_{c}\sim \phi_{0}(\ell/L)(\omega\widetilde{\beta})^{1/2}$ appear beyond a ``tricritical point'' (TP)
\begin{gather*}
\widetilde{\alpha}_{\mathrm{TP}} = \frac{\eta\Gamma(\pi/\phi_{0})^{2}(\ell/L)^{2}(D+1-2w)}{1-\lambda}\,,\\[5pt]
\widetilde{\beta}_{\mathrm{TP}} = \frac{(\pi/\phi_{0})^{2}(\ell/L)^{2}(D-w)^{2}}{w[1+(2\eta\Gamma)^{-1}(1-\lambda)^{2}]}\,.
\end{gather*}
This linearized  analysis predicts the positions of the ``phase boundaries'' S-SF and S-PF in {quantitative} agreement with  the numerical solution. The boundary SF-PF has  been obtained only numerically. The appearance of  spontaneous oscillations results from  the coupled motion of concentration and director orientation bands  due to both the convective active polar coupling ($\beta$) and the passive polar coupling ($w$) of director  and concentration. Upon increasing $\widetilde{\beta}$ the  oscillatory behavior becomes increasingly complex, but we have not been able to observe fully fledged chaos for reasonable values of $\widetilde{\beta}$.

We have studied the dynamical properties of thin films of active polar fluids and found a rich variety of complex behaviors which should be observable experimentally in polar active systems. Using microscopic models of motor-filament coupling, it was estimated in \cite{AATBLMCM06} that $\beta\gg\alpha$ in microtubules/kynesin mixtures, while $\beta\leq\alpha$ in actomyosin systems. This suggests that in-vitro microtubules/kynesin mixtures may be the best candidate for the observation of the oscillating bands predicted here. It should, however, be noted that filament  treadmilling also leads to terms with polar symmetry at the continuum level, where in this case $\beta$ would be proportional to the polymerization rate~\cite{JFJPKKFJ06}. The intriguing  possibility that our findings may be relevant to treadmilling acto-myosin systems and therefore have implications for lamellipodium dynamics will be explored elsewhere.

MCM and LG were supported on NSF  grants DMR-0305407 and DMR-0705105. MCM acknowledges  the hospitality of the Institut Curie and ESPCI  in Paris and the support of a  Rotschild-Yvette-Mayent sabbatical fellowship at Curie.  TBL acknowledges the hospitality of the Institut Curie in Paris and the support of the Royal Society and the EPSRC under grant EP/E065678/1. LG was supported on a Graduate Fellowship by the Syracuse Biomaterials Institute. Finally, we thank Jean-Francois Joanny and Jacques Prost for many useful discussions.

\end{document}